\documentstyle[12pt]{article}
 
\begin{document} 
\begin{center} 

Erratum to the paper 

Two-loop calculations for the propagators of gluonic currents.

By A.L. Kataev, N.V. Krasnikov, A.A. Pivovarov.

Published in Nucl. Phys. B198(1982)508. 
\end{center} 

Recently we have learned \cite{inf} that the result of our
computation of $\alpha_s$ corrections to the correlator of
gluonic
currents $G_{\mu\nu}G^{\mu\nu}$ in \cite{paper}
disagrees with \cite{new} by an amount proportional to the first 
coefficient of $\beta$ function.
We checked our files and discovered that we omitted the finite part of
one-loop diagram in our original presentation of results.
Corrected version is given below.

Scalar case (leading term is normalized to 1):
\[
\langle TG^2(x)G^2(0)\rangle
\rightarrow \Phi_B(Q)=Q^4\left({\mu^2\over Q^2}\right)^\epsilon
\left\{\left({1\over\epsilon}+B\right)
+
\left({\mu^2\over Q^2}\right)^\epsilon
{\alpha\over 4\pi}\left({\beta_0\over \epsilon^2}
+{D_{SG}\over \epsilon}\right)\right\},
\]
\[
\beta_0={11\over 3}N_c-{2\over 3}N_f, \quad
D_{SG}=
{7\over 6}N_c-{1\over 3}N_f,\quad
B=-1.
\]
\[\Phi_{SG}(Q)
=\alpha^2\left(1-2{\alpha\over 4\pi}{\beta_0\over \epsilon}\right)
\left(-Q^2{d\over d Q^2}\right)Q^{-4}\Phi_B(Q)=
\alpha^2\left(1+{\alpha\over 4\pi}(-2B\beta_0+2D_{SG})\right).
\]
The quantity $D_{SG}$ is a result of computation of two-loop diagrams.
It didn't change. The quantity $B$ is a finite part of one-loop diagram.
It was omitted in old presentation \cite{paper}.
At $B=0$ we reproduce our old results (Eq.~(15) of \cite{paper}).

At $B=-1$ the correct result (that is now in agreement with \cite{new})
reads
\[\Phi_{SG}(Q)=
\alpha^2(1+{\alpha\over 4\pi}({29\over 3}N_c-2N_f)).
\]
This is the result in G-scheme of renormalization at $\mu=Q$.
With logs included it reads
\[\Phi_{SG}(Q)=
\alpha^2(\mu)\left\{1+{\alpha\over 4\pi}({29\over 3}N_c-2N_f
+2\beta_0\ln\left(\mu^2\over Q^2\right))\right\}.
\]
To turn to $\overline{\rm MS}$-scheme one shifts 
$\mu^2\rightarrow \mu^2\exp(2)$ and finds
\[\Phi_{SG}^{\overline{\rm MS}}(Q)
=\alpha^2(\mu)\left\{1+{\alpha\over 4\pi}({73\over 3}N_c-{14\over 3}N_f
+2\beta_0\ln\left(\mu^2\over Q^2\right))\right\}.
\]

Thus Eq.~(16) of \cite{paper} now reads
\[
D_{SG}(1,\alpha)=\left\{
\begin{array}{ll}
{2\alpha^2\over \pi^2}\left({9\over 4\pi}\right)^2
\left[1+{\alpha\over \pi}
\left(\frac{23}{4}+\frac{32}{9}\right)\right],   &\left(G-{\rm
scheme}\right)\\
{2\alpha^2\over \pi^2}\left({9\over 4\pi}\right)^2
\left[1+{\alpha\over \pi}
\left(\frac{59}{4}+\frac{32}{9}\right)\right],   &\left(\overline{\rm MS}-{\rm
scheme}\right).
\end{array}\right.
\]

We have checked the pseudoscalar case as well
and found that the finite part of one-loop diagram was omitted also.
The corrected result now reads
(leading term is normalized to 1):
\[
\langle TG\tilde G(x)G\tilde G(0)\rangle
\rightarrow \tilde\Phi_B(Q)=Q^4\left({\mu^2\over Q^2}\right)^\epsilon
\left\{\left({1\over\epsilon}+B_{PG}\right)
+
\left({\mu^2\over Q^2}\right)^\epsilon
{\alpha\over 4\pi}\left({\beta_0\over \epsilon^2}
+{D_{PG}\over \epsilon}\right)\right\},
\]
\[
\beta_0={11\over 3}N_c-{2\over 3}N_f, \quad
D_{PG}=-{13\over 6}N_c+N_f, \quad
B_{PG}=-3.
\]
\[\Phi_{PG}(Q)
=\alpha^2\left(1-2{\alpha\over 4\pi}{\beta_0\over \epsilon}\right)
\left(-Q^2{d\over d Q^2}\right)Q^{-4}\tilde\Phi_B(Q)=
\alpha^2\left(1+{\alpha\over 4\pi}(-2B_{PG}\beta_0+2D_{PG})\right).
\]
The quantity $D_{PG}$ is a result of computation of two-loop diagrams.
It didn't change. The quantity $B_{PG}$ 
is a finite part of one-loop diagram.
It was omitted in old presentation in \cite{paper}.
At $B_{PG}=0$ we reproduce our old results.

At $B_{PG}=-3$ the corrected result
reads
\[
\Phi_{PG}(Q)=
\alpha^2(1+{\alpha\over 4\pi}({53\over 3}N_c-2N_f)).
\]
This is the result in G-scheme of renormalization at $\mu=Q$.
With logs included it reads
\[\Phi_{PG}(Q)=
\alpha^2(\mu)\left\{1+{\alpha\over 4\pi}({53\over 3}N_c-2N_f
+2\beta_0\ln\left(\mu^2\over Q^2\right))\right\}.
\]
To turn to $\overline{\rm MS}$-scheme one shifts 
$\mu^2\rightarrow \mu^2\exp(2)$ and finds
\[
\Phi_{PG}^{\overline{\rm MS}}(Q)
=\alpha^2(\mu)\left\{1+{\alpha\over 4\pi}({97\over 3}N_c-{14\over 3}N_f
+2\beta_0\ln\left(\mu^2\over Q^2\right))\right\}.
\]
Thus Eq.~(14) of \cite{paper} now reads
\[
D_{PG}(1,\alpha)=\left\{
\begin{array}{ll}
{2\alpha^2\over \pi^2}
\left[1+{\alpha\over 4\pi}
\left(47\right)\right],   &\left(G-{\rm
scheme}\right)\\
{2\alpha^2\over \pi^2}
\left[1+{\alpha\over 4\pi}
\left(83\right)\right],   &\left(\overline{\rm MS}-{\rm
scheme}\right).
\end{array}\right.
\]
Eq.~(1) of \cite{paper} becomes
\[
{\Lambda^2_{PG}\over \Lambda^2_{e^+e^-}}=
2.4 e^{-B_{PG}}=2.4 e^3=48.2, \quad
{\Lambda^2_{SG}\over \Lambda^2_{e^+e^-}}=10.4 e^{-B_{SG}}=10.4 e^1=28.3.
\]

Results for the pseudoscalar case given in \cite{pl}
should be changed accordingly.

\end{document}